\documentclass[aps,twocolumn]{revtex4}
\usepackage{graphicx}

\begin{document}

\title{Transition from tunneling to direct contact in tungsten nanojunctions}

\author{ A.~Halbritter, Sz.~Csonka and G.~Mih\'aly}
\affiliation{ Department of Physics, Institute of Physics,
Budapest University of Technology and Economics, 1111 Budapest,
Hungary}
\author{E.~Jurdik, O.Yu.~Kolesnychenko, O.I.~Shklyarevskii$^\dag$,
        S.~Speller and H.~van~Kempen}
\affiliation{NSRIM, University of Nijmegen, Toernooiveld 1, 6525
ED Nijmegen, the Netherlands}


\begin{abstract}
We apply the  mechanically controllable break junctions technique
to investigate the transition from tunneling to direct contact in
tungsten. This transition is quite different from that of other
metals and is determined by the local electronic properties of the
tungsten surface and the relief of the electrodes at the point of
their closest proximity. The conductance traces show a rich
variety of patterns from the avalanche-like jump to a mesoscopic
contact to the completely smooth transition between direct contact
and tunneling. Due to the occasional absence of an adhesive jump
the conductance of the contact can be continuously monitored at
ultra-small electrode separations. The conductance histograms of
tungsten are either featureless or show two distinct peaks related
to the sequential opening of spatially separated groups of
conductance channels. The role of surface states of tungsten and
their contribution to the junction conductance at sub-{\AA}ngstrom
electrode separations are discussed.
\end{abstract}


\maketitle

\section{Introduction}

The study of transition from tunneling to direct contact and
electrical transport through atomic-sized metallic conductors has
been the object of great attention during the last decade.
Different types of phenomena, related to both the quantum
character of transport and the atomic discreteness of the contact,
were observed in 3D nanoconstrictions produced by scanning
tunneling microscopes (STM) or mechanically controllable break
junctions (MCBJ) \cite{agrait_review}. In particular, conductance
measurements of breaking nanowires demonstrated a step-like
structure of conductance versus electrode separation traces
$G(z)$. For single-valence $s$-metals conductance plateaus are
close to the integer multiples of the quantum conductance unit
$G_0=2e^2/h$. However, simultaneous measurements of both the force
and the conductance for breaking gold nanowires demonstrated that
jumps between plateaus in the conductance staircase are always
correlated with relaxations of the mechanical force and,
therefore, with atomic rearrangements in the nanoconstriction
\cite{rubio}. Individual conductance curves $G(z)$ are inherently
irreproducible due to the different dynamical evolution of the
connective necks during the break. Therefore, analysis of
experimental data includes construction of conductance histograms
based on a large number of conductance traces. Peaks in the
conductance histograms are related to the statistically more
probable atomic configurations in the connective neck between the
electrodes \cite{krans}. For polyvalent metals the main (and
sometimes the only) feature in the conductance histograms is a
peak corresponding to one-atom point contact \cite{thesis}. The
conductance through such a contact is determined by a few
conductance channels intimately related to atomic orbitals
\cite{cuevas}. Transition from tunneling to single-atom contact
occurs in an avalanche-like way at an electrode separation of
$\sim 1.5$~\AA~due to the metallic adhesion forces \cite{sutton}.
This sudden jump in conductance precluded measurements of $G(z)$
at sub-{\AA}ngstrom distances between the electrodes for all
metals studied to date.

While measuring thermal expansion of MCBJ electrodes for different
materials, we took notice of the unusually high stability of
tungsten tunnel junctions at very close electrode separations
\cite{thermo}. In another study, transmission and scanning
electron microscopy images showed no evidence of connective neck
formation between tungsten wires \cite{correia}. Moreover,
measurements of adhesive forces between an atomically-defined
W(111) trimer tip and a Au(111) sample revealed no spontaneous
jump to contact \cite{cross}. These unusual properties of tungsten
nanocontacts motivated us to carry out further extensive
investigations.

In this paper we present our experiments with tungsten MCBJ. Our
aim was to examine the behavior of tungsten nanojunctions during
the transition between tunneling and direct contact and to what
extent this behavior is influenced by the unique mechanical and
electronic properties of tungsten. We show that very often the
adhesive jumps of conductance are absent and that the transition
to single atom contact is smooth. This permitted us to investigate
the junction at ultra-small electrode separations. By employing
the conductance histogram technique we determined the preferential
values of conductance during the fracture of the junction to be
about $1$~$G_0$ and $2$~$G_0$. Also, we measured the conductance
histograms of tantalum and molybdenum (as two nearest neighbors of
tungsten in the periodic table of elements) to provide additional
support for our interpretation.

\section{Experimental setup}

\begin{figure}
\includegraphics[width=0.5\columnwidth]{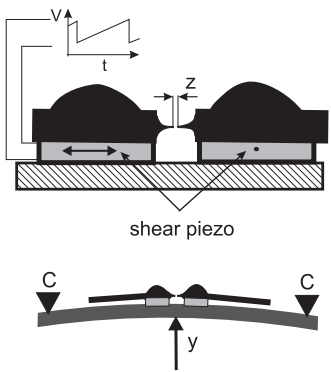}
\caption{Modified sample mounting. The tungsten wire is glued on
top of shear piezo-ceramic attached to a phosphor bronze bending
beam.  The distance between the electrodes is controlled by
vertically pushing the bending beam at the middle or by applying
voltage on the left-hand side shear piezo. All distances are
exaggerated for clarity.} \label{fig1}
\end{figure}

We employed the traditional MCBJ technique, described elsewhere
\cite{agrait_review}, with a modified sample mounting (Fig.~1).
This mounting includes two $6 \times 2.5 \times 1$~mm pieces of
shear piezo-ceramic (which gives horizontal displacement of its
surface when voltage is applied) glued at the center of a phosphor
bronze bending beam. A polycrystalline tungsten wire with a
diameter of 100~$\mu$m was attached to the top of the piezo's with
two small drops of hard epoxy (Stycast 850FT). The central section
of the wire was then electrochemically etched in a 25\% solution
of KOH down to $5$--$10$~$\mu$m at its thinnest part. The wire was
broken at $4.2$~K in an ultra-high vacuum environment by bending
the beam. The distance between the electrodes was fine tuned by
changing the deflection of the bending beam or applying voltage to
the left-hand side shear piezo. The relative displacement of the
electrodes was calibrated using the exponential part of $G(z)$
traces in the tunneling regime (assuming a work function for
tungsten $\phi \simeq 4.5$~eV) as well as by measuring field
emission resonance spectra \cite{rsi}. The contact conductance was
measured using a current-to-voltage converter with a gain of
0.1~V/$\mu$A. Conductance versus electrode separation traces
$G(z)$ were measured in a slow mode ($2$--$30$~points/s) using
Keithley 2182 nanovoltmeters in order to cover over 7 orders of
magnitude in the conductance during transition from tunneling to
direct contact. Conductance traces for building conductance
histograms were recorded with an AT-MIO-16XE-50 National
Instruments data acquisition board (sampling rate of
$20000$~points/s and resolution of $16$~bits). During this
acquisition a ramp voltage with a frequency of $5$--$50$~Hz was
applied to the shear piezo to establish a repeated fracture of the
junction.

The sample surfaces were characterized by scanning the electrodes
with the two shear piezo's in the constant current mode. We found
atomically flat parts up to $5$--$10$~nm in length alternating
with irregularities of $5$--$15$~\AA~ in height. It should be
noted that these STM-like measurements with two ``blunt''
electrodes are rather qualitative and the numbers presented are
only rough estimates.

We analyzed thoroughly more than $200$ conductance histograms
measured for $12$ different samples.

\section{Results}

\begin{figure}
\includegraphics[width=0.6\columnwidth]{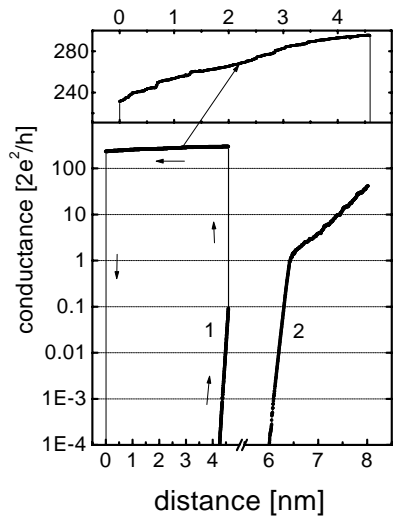}
\caption{Two extreme cases of $G(z)$ traces for W MCBJ. 1 --
avalanche-like transition from tunneling to a low-ohmic contact.
The upper panel shows the contact conductance in the process of
retraction. 2 -- transition from tunneling to direct contact
without spontaneous formation of an adhesive neck.} \label{fig2}
\end{figure}

We investigated the transition from the tunneling regime to direct
contact for tungsten MCBJ by performing conductance measurements
during slow, continuous approach or retraction of the electrodes.
These measurements revealed a rich variety of $G(z)$ traces. All
of them fall between two extreme cases shown in Fig.~\ref{fig2}.
Curve 1 demonstrates an avalanche-like transition from $G \approx
0.01$--$0.1$~$G_0$ to $100$--$1000$~$G_0$. This indicates the
formation of a mesoscopic contact with a cross-section of
$10$--$100$~nm$^2$ between two flat portions of the electrodes.
Subsequent retraction of the electrodes produced only small
changes in the contact conductance over a retreat distance of
$\gtrsim 4$~nm, which were followed by sudden disconnection (upper
panel in Fig.~\ref{fig2}). During $5$--$10$~minutes after
disconnection, the distance between the electrodes increased as
deduced from the transient decrease of the measured tunnel
current. After the full relaxation of the sample, the original
curve could be reproduced accurately, indicating that no
irreversible changes of the electrode relief occurred. The other
extreme case -- curve 2 in Fig.~\ref{fig2} -- demonstrates a
smooth transition from tunneling to direct contact with no sign of
spontaneous formation of an adhesive neck. The majority of $G(z)$
curves exhibited combination of smooth changes of the conductance
and sudden jumps with amplitudes ranging from $0.1G_0$ up to few
$G_0$.

\begin{figure}
\includegraphics[width=.6\columnwidth]{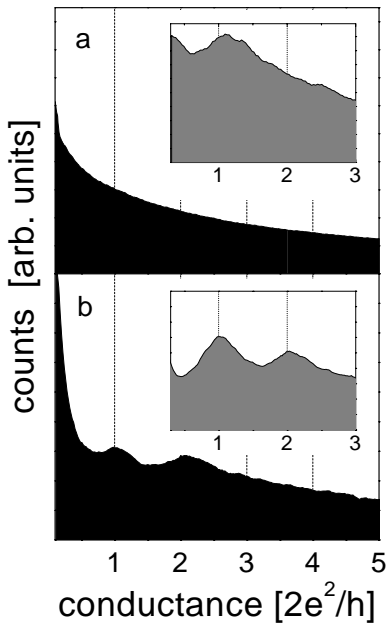}
\caption{Conductance histograms for W MCBJ based on $10000~G(z)$
traces. (a) -- featureless histogram. (b) -- histogram with peaks
close to integers of $2e^2/h$. Insets in (a) and (b): histograms
for the corresponding data sets after filtering out the
featureless curves.} \label{fig3}
\end{figure}

We also performed statistical analysis of conductance traces by
employing the conductance histogram technique. We found that
tungsten nanojunctions have a particular tendency to show
reproducible, almost identical conductance traces when indentation
of the electrodes is not sufficient. If proper care is taken and
in every cycle contacts of $G \gtrsim 30$--$40$~$G_0$ are formed,
the break occurs in a large variety of ways. In this latter case a
statistical approach to data analysis is justified. In sharp
contrast to other metals, the conductance histograms of tungsten
do not necessarily exhibit the same pattern when the contact site
is shifted by the right-hand side shear piezo or when the sample
is replaced. A major part of histograms demonstrated no
distinctive features [Fig.~\ref{fig3}(a)]. However, on many
occasions the histograms showed two peaks at respectively
$1$--$1.2$ and $2$--$2.2$~$G_0$ [Fig.~\ref{fig3}(b)], or only one
of those two.

\begin{figure}
\includegraphics[width=.6\columnwidth]{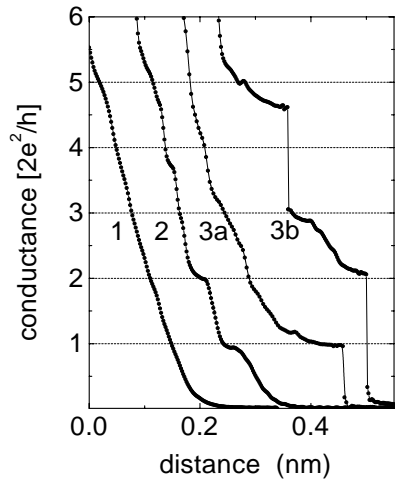}
\caption{Typical conductance traces for W. Curve $1$ --
featureless curve. Curve $2$ -- smooth transition to tunneling
with distinct plateaus. Curves $3$a and $3$b -- traces with sudden
jumps of conductance.} \label{fig4}
\end{figure}

In Fig.~\ref{fig4} four characteristic conductance traces are
presented. Curve 1 shows a smooth and completely featureless
transition from direct contact to tunneling. For a tungsten
junction this kind of transition is frequently present and gives
rise to the featureless background in the conductance histograms.
This background is suppressed when these featureless curves are
rejected from the data set by employing a computer filtering
algorithm \cite{note_filter}. Then, either peaks invisible in the
original histogram appear [inset in Fig.~\ref{fig3}(a)] or the
contrast of peaks is significantly improved [inset in
Fig.~\ref{fig3}(b)]. Peaks in the histogram naturally arise from
conductance traces with distinct features -- plateaus -- at about
$1G_0$ and/or $2G_0$ as demonstrated by curves 2, 3a and 3b in
Fig.~\ref{fig4}. While curve 2 shows a smooth transition with no
jumps, in the case of curves 3a and 3b clear jumps occur near
$1G_0$ and $2G_0$, respectively. We note here that conductance
traces like those represented by curve 2 are unique to tungsten
and were not observed in other metallic contacts.

\begin{figure}
\includegraphics[width=.6\columnwidth]{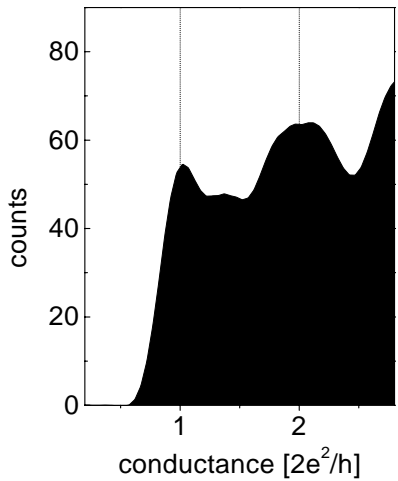}
\caption{Histogram for conductance values at which jumps with
amplitude $\Delta G > 0.7G_0$ occur.} \label{fig5}
\end{figure}

In our previous analysis we used an algorithm to select the traces
with long plateaus \cite{note_filter}. However, it is also
important to differentiate between the smooth curves and those
with sudden jumps. For this reason every trace was characterized
by the largest jump between neighboring data points in the
conductance interval $G/G_0 \in [0.3, 3.3]$. By combining these
two approaches the traces were classified into the three
categories presented in Fig.~\ref{fig4}. Using the  data set for
histogram in Fig~\ref{fig3}(b) we found that $\sim 30\%$ of traces
exhibited a smooth and featureless transition and roughly $10$\%
of curves show plateaus without sudden jumps. In the rest of the
traces clear conductance jumps were found. In order to quantify at
which conductance values  conductance jumps with $\Delta G_{max}
\geq 0.7 G_0$ are most likely to occur, we also built up
histograms for the position of these sudden transitions. The peaks
in the so-built histogram presented in Fig.~\ref{fig5} clearly
demonstrate that large conductance jumps are most probable at $G
\simeq 1G_0$ and $2G_0$.  We emphasize that the relative
occurrence of the three different types of conductance traces is
sample and site-dependent, and in certain data sets only traces
with smooth transitions were present.

\section{discussion}


Our experimental results for tungsten nanojunctions strikingly
differ from those obtained for other metals studied so far, that
exhibit conductance traces always falling into the same
(material-specific) pattern. Especially, the adhesive jump to
contact or the tensile overstretching followed by sudden
disruption have respectively been found to be an unavoidable
attendant of contact formation or breakage. This has precluded
measurements of conductance at small electrode separations. The
occasional absence of these processes is unique for tungsten
junctions and we attribute it to the specific mechanical
properties of this metal. In particular, this odd junction
stability at ultra-small electrode separations permitted us to
study the previously unaccessible regime of nanocontact
conductance.

The adhesive jump is governed by the the competition between the
adhesive and tensile forces \cite{taylor}. The tensile forces are
determined by the stiffness of the electrodes, which is composed
from the stiffness of the nanoscale junction itself and the
stiffness of the whole setup. The latter can have significant
importance in an atomic force microscope with a soft cantilever,
however it is not expected to play any role in an MCBJ due to the
rigid sample mounting. The stiffness of the junction depends on
the strength of the bonds in the material, but also the junction
geometry and the crystallographic orientation of the electrodes
have a crucial influence on it. Contacts with large opening angle
have large stiffness, while in contacts with small opening angle a
large amount of layers are involved in the elastic deformation and
thus the stiffness is reduced. The stiffness is also reduced, if
the contact surfaces are not perpendicular to the contact axis.
Then, the electrodes can bend like a cantilever and the elastic
behavior is defined by the smaller shear forces.

The elastic behavior of the material is basically determined by
the strength of the bonds between the atoms, therefore it is
similar at any length-scale and atomic scale systems are well
described by bulk elastic constants \cite{rubio}. The bulk elastic
modulus of tungsten ($E_{\mathrm{W}}=411$\,GPa) is outstandingly
large among all metals. (As a comparison the Youngs modulus of Au,
Pt, and Nb is $E_{\mathrm{Au}}=78$\,GPa,
$E_{\mathrm{Pt}}=168$\,GPa, and $E_{\mathrm{Nb}}=105$\,GPa,
respectively.) On the other hand, the junction geometry is not
controllable in MCBJ experiments. In previously studied metals the
conductance traces imply that the junction breaks through the
process of rupture. In this case the contact geometry is
``self-organized'' during the neck pulling, so the conditions for
the adhesive jumps are similar during each disconnection. Our
measurements imply that tungsten breaks in a more brittle way and
a large variety of contact geometries are established during the
fractures. This, together with the outstanding elastic modulus of
tungsten can explain our observations. In ``stiff'' contact
geometries the elastic forces can overcome the adhesion and
thereby enable a smooth transition between tunneling and direct
contact, while in softer contact geometries the adhesive jumps are
still present.

Our interpretation can be tested by comparing the mechanical
behavior during the break of further metals with different elastic
modulus. For this reason we performed measurements on two
neighbors of tungsten in the periodic system, molybdenum and
tantalum. These metals have similar electronic properties to
tungsten. The elastic modulus of molybdenum is relatively large
($E_{\mathrm{Mo}}=329$\,GPa), while tantalum is much softer
($E_{\mathrm{Ta}}=186$\,GPa). In agreement with that we found that
tantalum always breaks in a jump-like way, while in molybdenum a
part of the conductance traces showed a smooth transition.

To explain the avalanche-like motion of hundreds to thousands of
atoms in the extreme case presented in Fig.~\ref{fig2} as curve 1,
we consider the following three possibilities. The first one
corresponds to the motion of a number of atomic layers in the
direction normal to the electrode surfaces as assumed in the
calculations by Taylor \textit{et al.} \cite{taylor}. This
possibility would require an avalanche of a macroscopic amount of
metal. Such a situation can take place only for contacts that act
as a very soft spring and thus we do not consider it as the most
likely explanation. The second possibility, that would result in
the same behavior of the contact conductance, is the transition to
direct contact due to bending of the electrodes. This arrangement
is likely to occur when flat parts of the electrode surfaces are
not perpendicular to the electrode axis. Then, the component of
the adhesive force that is normal to the electrode axis causes
bending. During our experiments, we observed anomalously high
sensitivity of the tungsten MCBJ to acoustic vibrations, such as
human voice, that suggests a transverse, spring-like motion of the
electrodes. This observation is in favor of the ``bending model.''
As the last possibility, the contact between the electrodes may
emerge as a result of dislocation glide or homogeneous shear
motion of one of the electrodes \cite{agrait_review}. However, in
this case it is not possible to explain the reversible behavior of
contact conductance through many formation-breaking cycles.


The smooth conductance traces with plateaus at the first two
conductance quanta (e.g., curve 2 in Fig.~4) are very close to
those expected for conductance quantization in short constrictions
\cite{torres}. In spite of its attractiveness, we discarded this
effect as the possible origin of the peaks in our histograms.
Extremely small distances between the plateaus in conductance
traces of tungsten, $\Delta z \simeq 30$~pm, suggest a
constriction with an opening angle close to 90 degrees (an
orifice). In this case the conductance quantization must be
completely  suppressed \cite{torres}. The shape of these traces is
better explained by the sequential closing of conduction channels
due to different spatial distribution of the involved electronic
states.

For transition $d$-metals five conductance channels are expected
to contribute to the conductance of a single-atom contact
according to the number of valence orbitals. The transmission
values of these conductance channels were extensively studied for
single-atom niobium junctions by tight binding calculations
\cite{cuevas} and by measuring subgap structure in the
superconducting state \cite{niobium}. The calculations performed
for a simplified one-atom contact geometry revealed a single
dominant channel (as a result of the hybridization between the $s$
and $d_{z^2}$ orbitals), two medium-sized channels and two smaller
channels that yield together a net conductance of
$G=2$--$3$~$G_0$. Both the conductance and the transmission
coefficients showed a good agreement with the values determined
from the experiment. Such a detailed investigation of the
transmission eigenvalues is not available for the rest of the
$d$-metals, and therefore the conduction properties of monoatomic
contacts can only be deduced from the conductance histograms. For
all non-magnetic transition metals studied so far (Nb, V, Rh, Pd,
Ir, Pt) the histograms show a single well-defined peak centered in
the region of $G=1.5$--$2.5~G_0$, that is attributed to the
conductance through a single-atom contact
\cite{agrait_review,thesis}. To date, theoretical calculations
have not been performed for the conductance of a single-atom
tungsten contact, but its value is presumably in the same range of
$G \simeq 1.5$--$2.5~G_0$.

In tungsten the first peak in the histogram is situated at
$1$~$G_0$, which is too low to be attributed to a single atom
contact with all the channels open. Therefore, rather the second
peak at $2$~$G_0$ is interpreted as the conductance of a
monoatomic contact geometry, where the center atom is closely
bound to both electrodes. The peak at $1$~$G_0$ is assumably
connected to an arrangement where the electrodes are so much
separated that only a part of the electronic states have
noticeable overlap.

It is well known from STM studies that tungsten surfaces have long
protruding surface modes, which have a crucial role in achieving
atomic resolution. First principle calculations for the W(001)
surface demonstrated that the tunnel current is primarily
generated by the $5d_{z^2}$ dangling-bond surface states
\cite{ohnishi,chen}. At distances $>2$~\AA~from the nuclei the
charge density of the surface state is much higher than that of
the atomic $6s$-state \cite{post} and so the surface states play a
key role in the electron transport at small separations. The
conductance through a single channel formed by such surface states
can be a good candidate for explaining the conductance plateaus at
$1$~$G_0$.

The occasional absence of adhesive jump enables us to monitor the
transition between the two situations. However, the sequential
closing of conductance channels -- like curve 2 in Fig.~\ref{fig4}
-- may only occur for specific electrode configurations, which
explains the relatively low percentage of this type of conductance
traces.

\begin{figure}
\includegraphics[width=0.6\columnwidth]{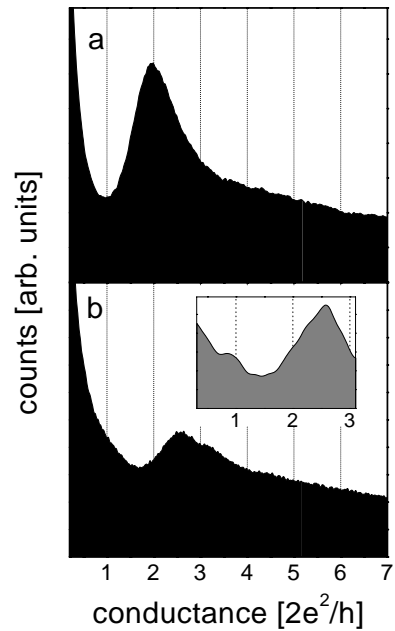}
\caption{Conductance histograms for tantalum (a) and molybdenum
(b) based on 20000 conductance traces. Inset in (b) shows a
histogram for selected 10\% of the curves, exhibiting plateaus
around 1$G_0$.} \label{fig6}
\end{figure}

The conductance histograms of tantalum and molybdenum were also
studied. These two neighbors of tungsten in the periodic system
have very similar electronic structure to W. Both for Ta and Mo
the histograms show a single, well-defined peak situated around
$2$~$G_0$ and $2.5$~$G_0$, respectively (see Fig.~\ref{fig6}).
This gives further support to the assumption that the conductance
through a single atom contact is in the range of $2$~$G_0$ for
tungsten as well. Though the electronic properties of Mo surfaces
are similar to tungsten \cite{molybdenum} the histograms show no
indication for a peak at $1$~$G_0$. In some occasions the computer
filtering of the original data set reveals conductance traces with
plateaus close to the conductance quantum, and accordingly a small
peak around 1~$G_0$ is recovered in the histogram (see inset in
Fig.~\ref{fig6}(b)).

\section{Conclusions}

In conclusion, our measurements have shown that the evolution of
conductance during the break of tungsten nanojunctions essentially
differs from the ususal behavior of metallic contacts. Due to the
large elastic modulus of W and the the non-ductile contact
fracture, the adhesive jumps in the conductance traces are
occasionally absent. Thank to these completely smooth transitions
from direct contact to tunneling, the closing of conductance
channels can be continuously monitored at ultra small electrode
separations. The study of conductance histograms has shown, that
the preferential values of conductance in tungsten junctions are
$1$\,$G_0$ and $2$\,$G_0$. The smooth conductance traces with
plateaus at these two values are explained by the sequential
closing of conductance channels due to their different spatial
distribution. The conduction through the long-protruding surface
states of tungsten might be responsible for the plateaus at
$1$\,$G_0$.

The limitations of MCBJ technique precluded us from drawing a more
exact and persuasive conclusion. The current data can greatly be
improved by simultaneous measurements of the conductance and the
forces in the course of electrode approach for contacts with a
well-defined tip/sample shape and orientation. To our opinion, the
details of the process of avalanche-like transition to direct
contact and subsequent break can be visualized with a high
resolution transmission electron microscope operating in UHV.

\section*{Acknowledgments}

The authors are grateful to A.J. Toonen, J. Hermsen and J.
Gerritsen for invaluable technical assistance. Part of this work
was supported by the Stichting voor Fundamenteel Onderzoek der
Materie (FOM) which is financially supported by the Nederlandse
Organisatie voor Wetenschappelijk Onderzoek (NWO). The Hungarian
Research Funds OTKA TO37451, TS040878, N31769 and a NWO grant for
Dutch-Hungarian cooperation are also acknowledged. O.I.S. wishes
to thank the NWO for a visitor's grant.



\begin{references}
\bibitem{note_filter} The integral
$\int(| \partial_z G(z)| + \varepsilon)^{-1} dz$, where
$\varepsilon > 0$ is a small number used to prevent singularities,
was calculated for each conductance trace within an interval
$G(z)/G_0 \in [0.3,3.3]$ and its value was used to select the
curves with distinct plateaus.
\end{references}
\end{document}